\documentclass[prx,twocolumn,floatfix,a4paper,superscriptaddress]{revtex4}

\usepackage{bm,color,amsmath,txfonts}
\usepackage{graphicx}
\usepackage{siunitx}
\usepackage{subfigure}
\usepackage{verbatim}
\usepackage{dcolumn}
\usepackage{bm}
\usepackage{epsf}
\usepackage{xcolor}
\usepackage{hyperref}
\usepackage{hhline}
\usepackage{float}
\usepackage{enumerate}
\usepackage{bbm}
\usepackage{lipsum}
\usepackage{mathrsfs}
\usepackage{amssymb}
\usepackage{ulem}

\begin{document}
	\title{Optomagnonic continuous-variable quantum teleportation enhanced by non-Gaussian distillation}
	\author{Zi-Xu Lu}
	\author{Xuan Zuo}
	\author{Zhi-Yuan Fan}
	\author{Jie Li}\thanks{jieli007@zju.edu.cn}
	\affiliation{Zhejiang Key Laboratory of Micro-Nano Quantum Chips and Quantum Control, School of Physics, and State Key Laboratory for Extreme Photonics and Instrumentation, Zhejiang University, Hangzhou 310027, China}

	\begin{abstract}
		The capability of magnons to coherently couple with various quantum systems makes them an ideal candidate to build hybrid quantum systems. The optomagnonic coupling is essential for constructing a hybrid magnonic quantum network, as the transmission of quantum information among remote quantum nodes must be accomplished using light rather than microwave field.  Here we provide an optomagnonic continuous-variable quantum teleportation protocol, which enables the transfer of an input optical state to a remote magnon mode. To overcome the currently relatively weak coupling in the experiment, we introduce non-Gaussian distillation operations to enhance the optomagnonic entanglement and thus the fidelity of the teleportation. An auxiliary microwave cavity is adopted to realize the non-Gaussian and displacement operations on magnons. We show that a series of optical states, such as coherent, single-photon, squeezed and cat states, can be teleported to the magnon mode. The work provides guidance for the experimental realization of magnonic quantum repeaters and quantum networks and a new route to prepare diverse magnonic quantum states exploiting the photon-to-magnon quantum teleportation.
		
	\end{abstract}
	
	
	\maketitle

	\section{Introduction}

	Quantum teleportation describes the transfer of an unknown input state onto a remote quantum system without the physical transfer of the carrier of quantum information~\cite{Bennett93}. 	It is an essential component for realizing many quantum protocols, such as quantum repeaters~\cite{Zoller98,Sangouard} and distributed quantum computing~\cite{Rau,Barz}. It was originally designed for and subsequently demonstrated in discrete-variable (DV) systems~\cite{Bou,Riebe,Sherson,Simon21}. Moreover, theories~\cite{Vaidman94,BK98} indicated that it can also been applied to continuous-variable (CV) systems with an infinite-dimension Hilbert space~\cite{Loock}.
	Relevant experiments were successfully demonstrated in, e.g., photonic~\cite{Furu98,Bowen03,TC03,Furu05} {and microwave~\cite{Deppe,Orcutt} systems} and atomic ensembles~\cite{Polzik06,Polzik13}.  
	Since entanglement is an essential element for quantum teleportation, a great deal of effort~\cite{Opatrny00,Cochrane02,Kitagawa06,Grangier07,Anno07,Yang09} was made to improve the performance of the teleportation utilizing non-Gaussian operations to distill the entanglement~\cite{Plenio02,J02}.

	In recent years, hybrid quantum systems based on magnons (the quanta of spin waves) in magnetic materials, e.g., yttrium iron garnet (YIG), have received great attention and achieved rapid development in both theory and experiment~\cite{Nakamuraapp,Yuan,Bauer,Zuo24}. The magnonic system shows two particularly compelling advantages: it exhibits great tunability, e.g., its resonance frequency can be continuously tuned within a wide range; and it demonstrates remarkable capability to coherently couple with diverse quantum systems, including microwave photons~\cite{Huebl,Nakamura14,Zhang14}, optical photons~\cite{NakamuraPRB,Nakamura16,Nakamura18,Zhang16,Haigh16}, phonons~\cite{ZhangSA,JieL,Davis21,Shen,Shen25,Dong}, and superconducting qubits~\cite{NakamuraSci15,NakamuraSA,NakamuraSci20,Xuda}. 
	In particular, the coupling between magnons with optical photons, e.g., of the whispering gallery modes (WGMs) of a YIG sphere, forms the system of optomagnonics~\cite{Nakamura16,Nakamura18,Zhang16,Haigh16}.  The optomagnonic coupling is vital for realizing quantum networks based on magnons~\cite{Jie21X}, since the transmission of quantum information among quantum nodes must be accomplished using optical photons rather than microwave photons (the latter suffers a huge transmission loss).  In the field of optomagnonics, a number of proposals have been offered to prepare magnonic non-Gaussian states~\cite{Bittencourt,Gao19,Qiong,Jiang21,Lu}, construct magnonic quantum networks~\cite{Jie21X,Zhao23}, test Bell inequalities~\cite{Xie22}, and realize the DV quantum teleportation and entanglement swapping~\cite{Fan}, to name a few. However, to date no protocol has been provided for realizing optomagnonic CV quantum teleportation. In comparison with the DV protocol~\cite{Fan}, which is, by its nature, probabilistic, the CV teleportation is deterministic. Moreover, the CV teleportation can transfer more information benefitting from the capability of the CV system to encode more information in infinite dimensional Hilbert spaces.

	Here, we present the first CV quantum teleportation protocol in optomagnonics and show how the protocol can be enhanced by non-Gaussian distillation operations. We adopt an optomagnonic system of a YIG sphere that is placed inside a microwave (MW) cavity. We first activate the optomagnonic Stokes scattering, namely, the magnon-induced Brillouin light scattering (BLS), to generate Gaussian entanglement between magnons and Stokes photons. We then apply non-Gaussian operations, i.e., subtracting a single photon and magnon, to distill and enhance the optomagnonic entanglement. With the distilled non-Gaussian entanglement shared between Alice and Bob, Alice mixes an initial input (optical) state with the Stokes field at a 50:50 beam splitter, then performs the homodyne detection (HD), and tells the measurement results (i.e., two optical quadratures) to Bob via classical communication. Depending on the results, Bob makes a displacement operation onto the magnon mode, preparing the magnon mode in the desired state corresponding to the initial state teleported from Alice to Bob. 
	We discuss a series of different input states, such as coherent, single-photon, squeezed, and cat states, and show that in all the cases the non-Gaussian operations can greatly enhance the optomagnonic entanglement and thereby improve the fidelity of the teleportation.

	\begin{figure*}[t]
		\includegraphics[width=\linewidth]{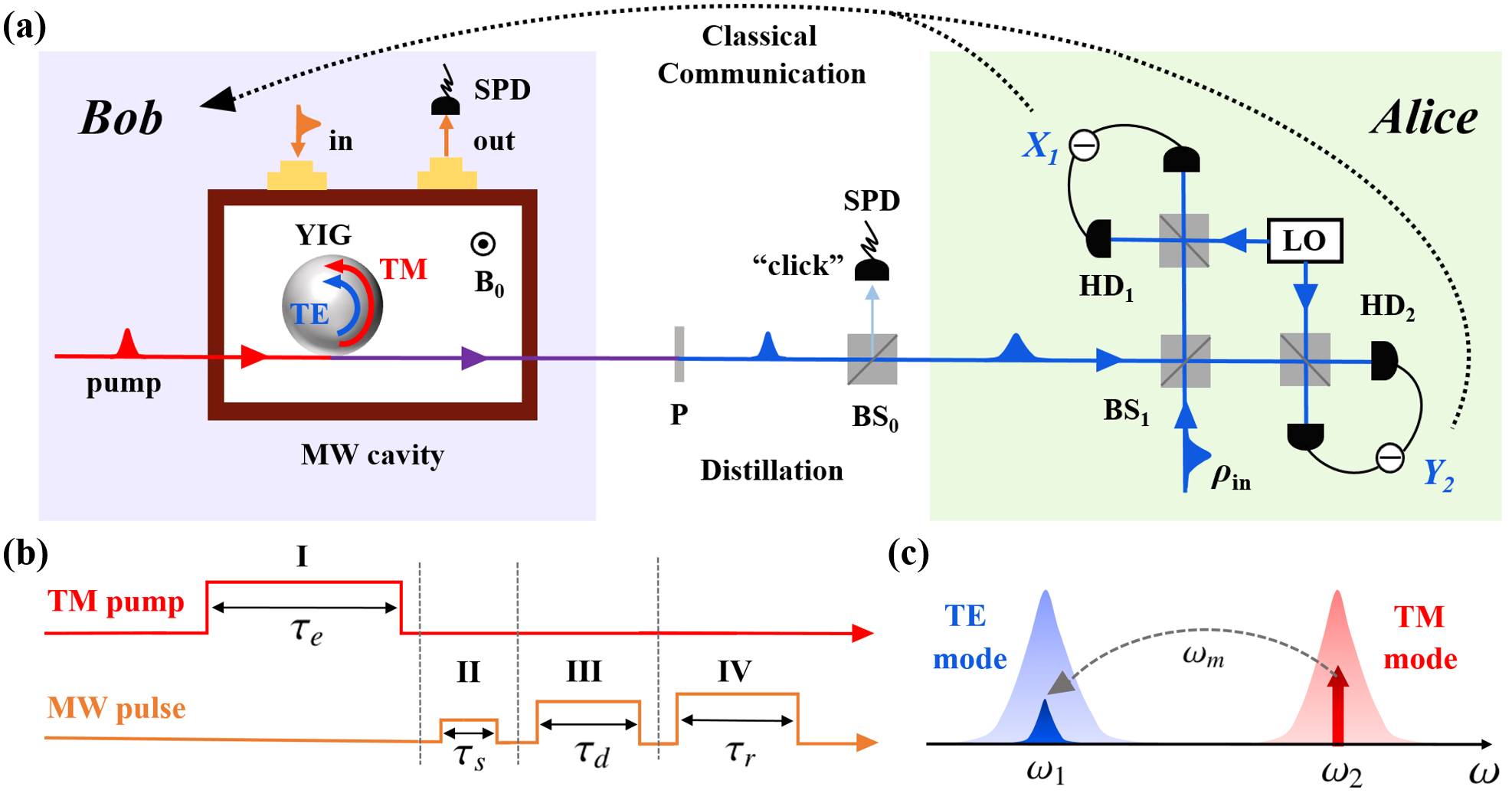}
		\caption{(a) Schematic diagram of optomagnonic CV quantum teleportation. A YIG sphere, supporting a magnon mode and two WGMs with different polarizations (i.e., the TM- and TE-polarized modes), is placed inside a MW cavity.  A strong optical pulse with duration $\tau_{e}$ is used to pump the TM-polarized WGM to activate the optomagnonic Stokes scattering, yielding entanglement between the magnon mode and TE-polarized Stokes field, which reaches a low-reflectivity beam splitter ($\mathrm{BS}_{0}$) after passing through a polarizer (P). 	A single photon is subtracted from the Stokes field when the single-photon detector (SPD) clicks at the reflection port of $\mathrm{BS}_{0}$. The single-photon-subtracted Stokes field is then mixed with an input optical state $\rho_{\rm in}$ at a 50:50 beam splitter ($\mathrm{BS}_{1}$).  On the other side, a single magnon is also subtracted from the magnon mode by sending a weak pulse with duration $\tau_{s}$ into the MW cavity and conditioned on the detection of a single MW photon in the output port.   
			Alice subsequently performs the homodyne detection (HD) on the two output fields of $\mathrm{BS}_{1}$ to measure a pair of quadratures $X_{1}$ and $Y_{2}$, and she then tells the results to Bob via classical communication. Based on the results ($X_{1}$, $Y_{2}$),  Bob implements a displacement operation onto the magnon mode by sending another pulse with duration $\tau_{d}$ into the MW cavity, which completes the teleportation. The final magnon state can be read out by sending one last MW pulse with duration $\tau_{r}$.
			(b) Time sequence of the optical and MW pulses used in the protocol.  The related description is provided in (a). To minimize the influence of dissipation on the magnon state, the total time of all the pulses is assumed to satisfy $\tau_{\rm total}= \tau_{e} + \tau_{s} + \tau_{d} + \tau_{r} \ll \kappa_{m}^{-1}$.
			(c) Magnon-induced Stokes scattering used for creating the optomagnonic entanglement, i.e., a TMSV state. }
		\label{fig1}
	\end{figure*}

	The paper is organized as follows. In Sec.~\ref{II}, we introduce the system and outline the key steps for realizing the optomagnonic CV teleportation. In Sec.~\ref{III}, we explicitly show how the optomagnonic Gaussin entanglement can be created and distilled with non-Gaussian operations. In Sec.~\ref{IV}, we complete the teleportation by performing the HD and displacement operation, and present the results of the teleportation fidelity for different input states. Finally, we discuss and summarize the work in Sec.~\ref{V}.

	\section{The system and protocol}\label{II}
	
	The proposed photon-to-magnon quantum teleportation employs a typical optomagnonic system, i.e., a YIG sphere, placed inside a MW cavity, as depicted in Fig.~\ref{fig1}(a). The YIG sphere supports a magnetostatic mode~\cite{Gurevich}, e.g., the Kittel mode, and two optical WGMs with different polarizations. In this system, the optomagnonic interaction, i.e., the magnon-induced BLS, promises that photons in a WGM are scattered by lower-frequency magnons (typically in gigahertz), yielding two optical sidebands whose frequencies deviate from the resonance of the WGM by the magnon frequency. The scattering probability is maximized when the triple-resonance condition is satisfied~\cite{Nakamura16,Zhang16,Haigh16}, i.e., the scattered photons enter another WGM of the YIG sphere. The BLS exhibits a pronounced asymmetry in the Stokes and anti-Stokes scatterings, resulting from the selection rule imposed by the conservation of the angular momenta of the WGM photons and magnons~\cite{Sharma17,PAP17,Nakamuranjp,Haigh18}. 
	In addition, the selection rule causes different polarizations of the two WGMs, e.g., the transverse-magnetic (TM) and transverse-electric (TE) polarized WGMs denoted by different colors in Fig.~\ref{fig1}(a). The YIG sphere placed inside a MW cavity allows magnons to couple with MW cavity photons via the magnetic dipole interaction. 
	
	The Hamiltonian of the whole system is given by
	\begin{equation}
		H=H_{0}+H_{1}+H_{2},
	\end{equation}
	where $H_{0}/\hbar=\omega_{m}m^{\dagger} m$ is the free Hamiltonian of the magnon mode, with $m$ ($m^{\dagger}$) being the annihilation (creation) operator and $\omega_{m}$ being the resonance frequency of the magnon mode, which is tunable by varying the bias magnetic field $B_0$. 	    
	The Hamiltonian $H_{1}$ is the sum of the free Hamiltonian of the two WGMs, the optomagnonic interaction Hamiltonian, and the driving Hamiltonian, i.e.~\cite{Jie21X}
	\begin{equation}\label{optmag}
		\begin{split}	
			H_{1}/\hbar= &\omega_{1}a^{\dagger}_{1}a_{1}+\omega_{2}a^{\dagger}_{2}a_{2} +i E_{j}\left(a_{j}^{\dagger}e^{-i\omega_{p_{j}}t}- {\rm H.c.} \right) \\
			&+ g_{0} \left(a_{1}^{\dagger} a_{2} m^{\dagger} +a_{1} a_{2}^{\dagger} m \right),\\
		\end{split}
	\end{equation}
	where $a_{j}$ ($a_{j}^{\dagger}$) is the annihilation (creation) operator of the $j$th WGM, $j=1,2$, and $\omega_{j}$ are their resonance frequencies, satisfying the triple-resonance condition $|\omega_{2}-\omega_{1}|=\omega_{m}$. Current optomagnonic experiments suffer from a weak bare coupling $g_{0}$~\cite{Nakamura16,Nakamura18,Zhang16,Haigh16}, and an intense drive is usually adopted to greatly enhance the effective optomagnonic coupling strength $G$, with the corresponding coupling strength $E_{j}=\sqrt{2P_{j}\kappa_{j}/\hbar\omega_{p_{j}}}$ between the $j$th WGM  (with decay rate $\kappa_{j}$) and the drive field with frequency $\omega_{p_{j}}$ and power $P_{j}$. {However, we note that the bare coupling $g_{0}$ can be significantly improved by reducing the mode volume and increasing the mode overlap, e.g., $g_{0}$ can reach the kHz level for a micron-size YIG disk or ring microcavity~\cite{VI}. This will greatly reduce the power needed to get the MHz level effective coupling $G$ used later in Secs.~\ref{III} and~\ref{IV}}.  In what follows, we assume that the mode $a_{1}$ ($a_{2}$) represents the TE (TM)-polarized mode of a certain WGM orbit and, for simplicity, that the magnon-induced BLS occurs only between the TM and TE modes with the same WGM index, i.e., the orbital angular momentum of the WGM photons is conserved~\cite{Nakamura16,Nakamuranjp}.  In this case, the frequency of the TM mode is higher than that of the TE mode, $\omega_{\rm TM} > \omega_{\rm TE}$, due to the geometrical birefringence~\cite{Nakamura16}.

	The Hamiltonian $H_{2}$ describes the free Hamiltonian of the MW cavity, the cavity-magnon interaction Hamiltonian, and the MW driving Hamiltonian, given by
	\begin{equation}\label{cmBS}
		H_{2}/\hbar= \omega_{c}c^{\dagger}c+ g_{c}\left(cm^{\dagger}+c^{\dagger}m \right) +iE_{d} \left(c^{\dagger}e^{-i\omega_{d}t}- {\rm H.c.} \right),\\
	\end{equation}
	where $c$ ($c^{\dagger}$) denotes the annihilation (creation) operator of the MW cavity mode, with the resonance frequency $\omega_{c}$, and $g_{c}$ is the cavity-magnon coupling strength. Here, $E_{d}=\sqrt{2P_{d}\kappa_{c}/\hbar\omega_{d}}$ is the coupling strength between the cavity mode (with decay rate $\kappa_{c}$) and the MW drive field with frequency $\omega_{d}$ and power $P_{d}$.

	Following the Braunstein-Kimble protocol for CV quantum teleportation~\cite{BK98}, our distillation-enhanced optomagnonic CV teleportation contains four steps (cf. Fig.~\ref{fig1}(b)): $i$) Creating the photon-magnon entanglement; 
	$ii$) Distilling the entanglement with non-Gaussian operations, explicitly the single-magnon (photon) subtraction;  
	$iii$) Performing the HD and magnon displacement operation;  
	$iv$) Reading out the magnon state.  
	In the following sections, we show in detail how the above steps are realized.

	\section{Creating and distilling optomagnonic entanglement}\label{III}
	
	The optomagnonic entanglement is an essential element for realizing the teleportation. Since current optomagnonic experiments suffer from a relatively weak optomagnonic coupling~\cite{Nakamura16,Nakamura18,Zhang16,Haigh16}, which constrains the degree of the entanglement and thus the fidelity of the teleportation. To resolve this issue, we introduce an entanglement distillation procedure in our protocol, specifically non-Gaussian operations to enhance the entanglement. 
	
	To generate the optomagnonic entanglement, we send a strong optical pulse with duration $\tau_{e}$ to drive the TM-polarized WGM.   The subsequent optomagnonic interaction is governed by the Hamiltonian $H_{\rm OM}=H_{0}+H_{1}$, which, in the frame rotating at the pump frequency $\omega_{p_{j}}$, is given by
	\begin{equation}
		\begin{split}
			H_{\rm OM}/\hbar=&\Delta_{1}a^{\dagger}_{1}a_{1}+\Delta_{2}a^{\dagger}_{2}a_{2}+\omega_{m}m^{\dagger}m \\ &+g_{0}\left(a_{1}^{\dagger} a_{2} m^{\dagger}+a_{1} a_{2}^{\dagger} m \right) +i E_{j}\left(a_{j}^{\dagger}-a_{j} \right),\\
		\end{split}
	\end{equation}
	where $\Delta_{j}=\omega_{j}-\omega_{p_{j}}$ ($j=1,2$).  We consider that the WGM $a_{2}$ is resonantly driven, i.e., $\Delta_{2}=0$ and $\Delta_{1}=-\omega_{m}$ (Fig.~\ref{fig1}(c)), and further treat classically the strong drive field as a number~\cite{Jie21X}, i.e., assuming $\alpha_{2}\equiv\langle a_{2}\rangle=E_{2}/\kappa_{2}$, with $\kappa_{2}$ being the decay rate of the WGM $a_{2}$. This leads to the following linearized Hamiltonian in the interaction picture
	\begin{equation}\label{TMSHal}
		H_{\rm TMS}/\hbar=G_{1}\left(a_{1}^{\dagger}m^{\dagger}+a_{1}m \right),
	\end{equation}
	with $G_{1}=g_{0}\alpha_{2}$ being the pump-enhanced optomagnonic coupling strength. Equation~\eqref{TMSHal} describes a two-mode squeezing interaction between the TE-polarized WGM and the magnon mode, corresponding to the optomagnonic Stokes scattering (Fig.~\ref{fig1}(c)), where a TM-polarized pump photon converts into a TE-polarized Stokes photon by emitting a magnon. Assuming that $\tau_{e}\ll\kappa_{m}^{-1}$, the magnon dissipation during the pump pulse is negligible~\cite{Jie21X,TJK14}. This yields the following quantum Langevin equations (QLEs) during the pulse
	\begin{equation}
		\begin{aligned}
			&\dot{a}_{1}=-\kappa_{1}a_{1}-iG_{1}m^{\dagger}+\sqrt{2\kappa_{1}}a_{1}^{\rm in},\\
			&\dot{m}=-iG_{1}a_{1}^{\dagger},\\
		\end{aligned}
	\end{equation}
	where $\kappa_{1}$ ($a_{1}^{\rm in}$) is the decay rate (input field) of the WGM $a_{1}$. We consider a weak coupling $G_{1}\ll\kappa_{1}$, which is the case in the experiment due to a large decay rate of the WGM of a YIG sphere~\cite{Nakamura16,Nakamura18,Zhang16,Haigh16}. This allows for the adiabatic elimination of the WGM, yielding $a_{1}\simeq\kappa_{1}^{-1}(-iG_{1}m^{\dagger}+\sqrt{2\kappa_{1}}a_{1}^{\rm in})$. Using the input-output relation $a_{1}^{\rm out}=\sqrt{2\kappa_{1}}a_{1}-a_{1}^{\rm in}$, we obtain 
	\begin{equation}\label{a1-m}
		\begin{aligned}
			&a_{1}^{\rm out}=-i\sqrt{2\mathcal{G}_{1}}m^{\dagger}+a_{1}^{\rm in},\\
			&\dot{m}=\mathcal{G}_{1}m-i\sqrt{2\mathcal{G}_{1}}a_{1}^{\rm in \dagger},\\
		\end{aligned}
	\end{equation}
	with $\mathcal{G}_{1}\equiv G_{1}^{2}/\kappa_{1}$. We further define a set of normalized temporal modes under the driving pulse with duration $\tau_{e}$~\cite{Hofer11}
	\begin{equation}
		\begin{aligned}
			&A_{1}^{\rm in}(\tau_{e})=i\sqrt{\dfrac{2\mathcal{G}_{1}}{1-e^{-2\mathcal{G}_{1}\tau_{e}}}}\int_{0}^{\tau_{e}}e^{-\mathcal{G}_{1}s}a_{1}^{\rm in}(s)ds,\\
			&A_{1}^{\rm out}(\tau_{e})=i\sqrt{\dfrac{2\mathcal{G}_{1}}{e^{2\mathcal{G}_{1}\tau_{e}}-1}}\int_{0}^{\tau_{e}}e^{\mathcal{G}_{1}s}a_{1}^{\rm out}(s)ds,\\
		\end{aligned}
	\end{equation}
	which satisfy the canonical commutation relation $[A_{1}^{k},A_{1}^{k\dagger}]=1$ ($k=$ in, out). By integrating Eq.~\eqref{a1-m}, we achieve
	\begin{equation}\label{aaaaa}
		\begin{aligned}
			&A_{1}^{\rm out}(\tau_{e})=\sqrt{e^{2\mathcal{G}_{1}\tau_{e}}-1}m^{\dagger}(0)+e^{\mathcal{G}_{1}\tau_{e}}A_{1}^{\rm in}(\tau_{e}),\\
			&m(\tau_{e})=e^{\mathcal{G}_{1}\tau_{e}}m(0)+\sqrt{e^{2\mathcal{G}_{1}\tau_{e}}-1}A_{1}^{\rm in\dagger}(\tau_{e}).
		\end{aligned}
	\end{equation}
	Equation~\eqref{aaaaa} allows us to extract a propagator $U(\tau_{e})$ satisfying $A_{1}^{\rm out}(\tau_{e})=U^{\dagger}(\tau_{e})A_{1}^{\rm in}(\tau_{e})U(\tau_{e})$ and $m({\tau_{e}})=U^{\dagger}(\tau_{e})m(0)U(\tau_{e})$, given by~\cite{TJK14}
	\begin{equation}\label{eq:13}
		U(\tau_{e})=e^{\tanh r\,A_{1}^{\rm in\dagger}m^{\dagger}}\cosh r^{-(1+A_{1}^{\rm in\dagger}A_{1}^{\rm in}+m^{\dagger}m
			)}e^{\tanh r\,A_{1}^{\rm in}m},
	\end{equation}
	where the squeezing parameter $r$ is defined via $\cosh r=e^{\mathcal{G}_{1}\tau_{e}}$ and $\tanh r=\sqrt{1-e^{-2\mathcal{G}_{1}\tau_{e}}}$, implying that the squeezing $r$ grows with the product $\mathcal{G}_{1}\tau_{e}$ of the pulse strength and duration.
	
	In our protocol, the MW cavity and the YIG sphere are assumed to be in a dilution refrigerator maintained at a low temperature, e.g., tens of mK. For the magnon mode with a typical frequency in gigahertz~\cite{Nakamura14,Zhang14}, it is essentially in the vacuum state.  Therefore, for the initial state $|0\rangle_{m}|0\rangle_{1}$, the system, at the end of the optical pulse, is prepared in the state
	\begin{equation}\label{bbbb}
		|\varphi(\tau_{e})\rangle_{m,1}\!=\!U(\tau_{e})|0\rangle_{m}|0\rangle_{1}\!=\!\sqrt{1-\tanh^{2}\!r}\sum_{n=0}^{\infty}(\tanh r)^{n}|n,n\rangle_{m,1},
	\end{equation}
	which is a two-mode squeezed vacuum (TMSV) state of the magnon mode and the output (Stokes) field of the TE-polarized WGM. The relatively weak coupling strength $G_{1}$ in the experiment yields a small squeezing $r$, resulting in an unsatisfactory fidelity of the quantum teleportation~\cite{Plenio02,J02}. Therefore, in what follows, we shall distill the entanglement by subtracting a single magnon and photon from the above TMSV state~\cite{Opatrny00,Cochrane02,Kitagawa06,Grangier07,Anno07,Yang09}.   Without loss of generality and to calculate in an orderly manner, we first subtract a single magnon and then a single photon.

	Since the magnon state is a solid state and one cannot directly realize the subtraction, we utilize the MW cavity and its linear excitation-exchange (beam-splitter) interaction with the magnon mode, as seen in Eq.~\eqref{cmBS}. Specifically, a weak and short MW pulse with duration $\tau_{s}$ is sent into the cavity, and the Hamiltonian of the cavity-magnon system is $H_{\rm CM}=H_{0}+H_{2}$. Again, we assume $\tau_{s}\ll\kappa_{m}^{-1}$ to neglect the magnon dissipation during the subtraction pulse. In the case of the cavity, the magnon mode, and the pulse being resonant, the corresponding QLEs, in the frame rotating at the pulse frequency, read
	\begin{equation}
		\begin{aligned}
			&\dot{c}=-\kappa_{c}c-ig_{c}m+\sqrt{2\kappa_{c}}c^{\rm in},\\
			&\dot{m}=-ig_{c}c,\\
		\end{aligned}
	\end{equation}
	where $c^{\rm in}$ denotes the input field of the MW cavity. We consider a relatively large cavity decay rate, such that the weak-coupling condition $g_{c}\ll\kappa_{c}$ is also fulfilled, which allows us to adiabatically eliminate the MW cavity and obtain $c \simeq \kappa_{c}^{-1} (-ig_{c}m+\sqrt{2\kappa_{c}}c^{\rm in} )$. By further using the input-output relation $c^{\rm out}=\sqrt{2\kappa_{c}}c-c^{\rm in}$, defining the temporal modes
	\begin{equation}
		\begin{aligned}
			&C^{\rm in}(\tau_{s})=\sqrt{\dfrac{2\mathcal{G}_{c}}{e^{2\mathcal{G}_{c}\tau_{s}-1}}}\int_{0}^{\tau_{s}}e^{\mathcal{G}_{c}s}c^{\rm in}(s)ds,\\
			&C^{\rm out}(\tau_{s})=\sqrt{\dfrac{2\mathcal{G}_{c}}{1-e^{-2\mathcal{G}_{c}\tau_{s}}}}\int_{0}^{\tau_{s}}e^{-\mathcal{G}_{c}s}c^{\rm out}(s)ds,\\
		\end{aligned}
	\end{equation}
	with $\mathcal{G}_{c} \equiv g_{c}^2/\kappa_c$, and following the same procedures introduced previously, we accordingly extract the propagator
	\begin{equation}
		L(\tau_{s})=e^{-i\tan\theta\,C^{\rm in\dagger}m}\cos\theta^{-(C^{\rm in\dagger}C^{\rm in}-m^{\dagger}m)
		}e^{i\tan\theta\,C^{\rm in}m^{\dagger}},
	\end{equation}
	which corresponds to the interaction at the presence of the MW pulse. Here, we have defined $\cos\theta=e^{-\mathcal{G}_{c}\tau_{s}}$ and $\tan\theta=\sqrt{e^{2\mathcal{G}_{c}\tau_{s}}-1}$. For the initial state $|0\rangle_{c}|\varphi(\tau_{e})\rangle_{m,1}$, i.e., the MW cavity is in vacuum and the magnon and output Stokes field are in the state~\eqref{bbbb}, the whole system, at the end of the pulse, is prepared in the state (unnormalized)
	\begin{equation}\label{sub}
		\begin{split}
			|\psi(\tau_{s})\rangle_{c,m,1}
			&=\sum_{n=0}^{\infty}\dfrac{\big(-i\tan\theta \big)^{n}}{n!}\left(C^{\rm in\dagger}m \right)^{n}|0\rangle_{c}|\varphi'(\tau_{e})\rangle_{m,1}\\
			&\approx|0\rangle_{c}|\varphi'(\tau_{e})\rangle_{m,1}-i\tan\theta\,|1\rangle_{c}\left(m|\varphi'(\tau_{e})\rangle_{m,1}\right),
		\end{split}
	\end{equation}
	where
	\begin{equation}\label{mmmm}
		|\varphi'(\tau_{e})\rangle_{m,1}=\sqrt{1-(\tanh r\cos\theta)^{2}}\sum_{n=0}^{\infty}(\tanh r\cos\theta)^{n}|n,n\rangle_{m,1}
	\end{equation}
	signifies the change of the optomagnonic initial state $|\varphi(\tau_{e})\rangle_{m,1}$ due to the MW pulse.  In getting the last line of Eq.~\eqref{sub}, we have omitted higher-order terms with $n \ge 2$, which is a good approximation when $\tan^{2}\theta\ll 1$, i.e., $\mathcal{G}_{c}\tau_{s}\ll 1$. Equation~\eqref{sub} indicates that the subtraction of a single magnon $m|\varphi'(\tau_{e})\rangle_{m,1}$ can be achieved conditioned that a single MW photon in the cavity output field is detected, and the successful probability is approximately $\tan^{2}\theta$. {Recent microwave quantum experiments have already achieved a relatively high single-photon detection efficiency, e.g., 71\% in Ref.~\cite{Besse}  and 84\% in    Ref.~\cite{Kono}.}  It is worth noting that, as revealed by Eq.~\eqref{mmmm}, the squeezing parameter of the TMSV state is slightly reduced, because $\tanh r\rightarrow \tanh r \cos\theta=e^{-\mathcal{G}_{c}\tau_{s}} \tanh r$.

	Soon after a single magnon being subtracted, we subtract a single photon from the Stokes output field. This can be realized by placing a low-reflectivity $\mathrm{BS}_{0}$ before the Stokes field entering the homodyne measurement stage (Fig.~\ref{fig1}(a)).  The other input of $\mathrm{BS}_{0}$ is vacuum, such that if a single photon is detected at the reflection port, the subtraction of a single photon from the Stokes field is completed. This corresponds to the joint state of the magnon mode and the Stokes output field being a non-Gaussian entangled state 
	\begin{equation}\label{nonGaus}
		A_{1}^{\rm out}m\,|\varphi'(\tau_{e})\rangle_{m,1},
	\end{equation}
	where $A_{1}^{\rm out}$ denotes the annihilation operator of the output temporal mode. Note that since the optomagnonic scattering probability is relatively low, the output field contains TE-polarized Stokes photons as well as unscattered TM-polarized pump photons. A polarizer is placed before $\mathrm{BS}_{0}$, such that only Stokes photons enter the subsequent measurements.

	\begin{figure}[t]
		\hskip-0.4cm \includegraphics[width=0.95\linewidth]{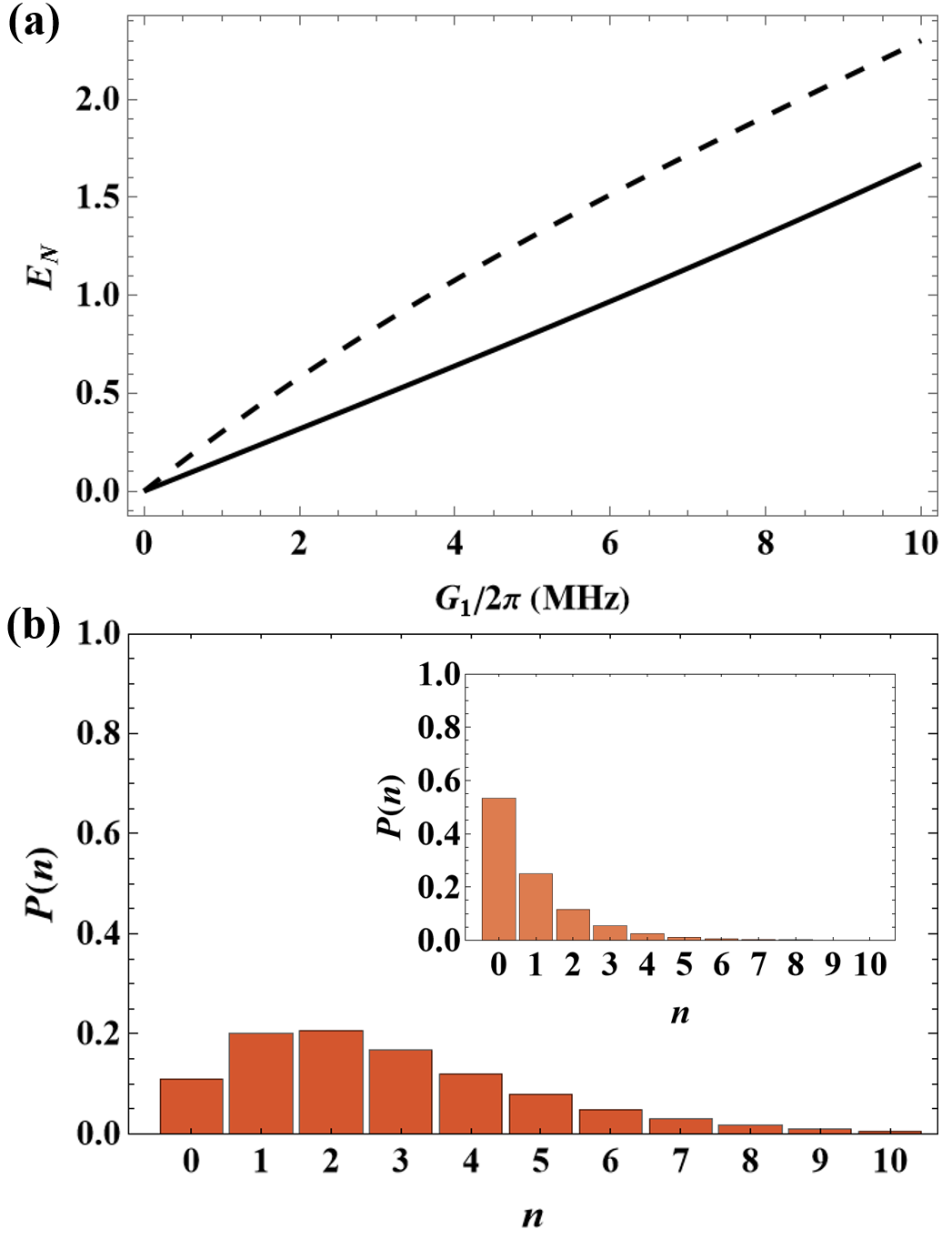}
		\caption{(a) Entanglement $E_{N}$ of non-Gaussian entangled state~\eqref{nonGaus} (dashed line) and TMSV state~\eqref{bbbb} (solid line) versus effective optomagnonic coupling strength $G_{1}$. (b) Magnon-photon joint number distribution of non-Gaussian state~\eqref{nonGaus} and TMSV state~\eqref{bbbb} (inset). We use experimentally feasible parameters: $\kappa_{m}/2\pi=0.5$ MHz~\cite{Shen25}, $\tau_{e}=50$ ns, $\kappa_{1}/2\pi=100$ MHz, $g_{c}/2\pi=4$ MHz, $\kappa_{c}/2\pi=40$ MHz, and $\tau_{s}=4$ ns, which yield $\mathcal{G}_{c}\tau_{s}\approx 0.01$.}
		\label{fig2}
	\end{figure}

	To characterize the entanglement of the non-Gaussian state~\eqref{nonGaus}, we adopt the logarithmic negativity $E_{N}=\ln[1+2 \mathcal{N} (\rho)]$~\cite{Vidal02}, where $\mathcal{N} (\rho)$ is the absolute value of the sum of the negative eigenvalues of the state $\rho$ after the partial transpose. For the TMSV state $|\varphi(\tau_{e})\rangle_{m,1}$, $E_{N}$ increases with the squeezing parameter $r$ via $E_{N}=2r$. Following Refs.~\cite{Cochrane02,Kitagawa06}, we calculate the logarithmic negativity of the non-Gaussian entangled state~\eqref{nonGaus}, i.e.,
	\begin{equation}
		E_{N}=\ln \dfrac{\big(1+\tanh r\cos\theta\big)^{3}}{\big(1+\tanh^{2}\!r\,\cos^{2}\!\theta\big)\big(1-\tanh r\cos\theta\big)}.
	\end{equation}
	In Fig.~\ref{fig2}(a), we compare the entanglement $E_{N}$ of both the Gaussian state~\eqref{bbbb} and non-Gaussian state~\eqref{nonGaus} versus the optomagnonic coupling $G_{1}$. A larger $G_{1}$ gives a larger squeezing parameter $r$ and thus a greater entanglement $E_{N}$.	Clearly, the non-Gaussian entanglement is higher than the Gaussian entanglement in the full range of $G_{1}$, which manifests that the non-Gaussian photon/magnon subtraction operation can distill the Gaussian entanglement. In Fig.~\ref{fig2}(b), we show the magnon-photon joint number distribution of the entangled states $|\varphi(\tau_{e})\rangle_{m,1}$ and $A_{1}^{\rm out}m\,|\varphi'(\tau_{e})\rangle_{m,1}$.
	It shows that the non-Gaussian subtraction operation significantly changes the magnon/photon number distribution, leading the non-Gaussian entanglement to have more contribution from the correlations of higher-order number states~\cite{Cochrane02}, which essentially causes the distillation of the entanglement.

	\section{Optomagnonic quantum teleportation}\label{IV}

	After preparing a non-Gaussian entangled state shared between Alice and Bob, we explicitly show below how to complete the teleportation of a (quantum) state from an input optical pulse to the magnon mode by means of the HD and classical communication.
	
	An optical pulse carrying an initial state $\rho_{\rm in}$  is mixed with the TE-polarized Stokes field at a 50:50 $\mathrm{BS}_{1}$ (Fig.~\ref{fig1}(a)).  Alice then performs the HD on the two output fields of $\mathrm{BS}_{1}$ to measure two quadratures $X_{1}$ and $Y_{2}$ (here the subscript distinguishes two output ports of $\mathrm{BS}_{1}$), and tells the measurement results ($X_{1}$, $Y_{2}$) to Bob via classical communication.  Note that the optical pulse and the local oscillator (LO) for the HD must be mode matched to the subtracted TE-polarized Stokes field, i.e., they possess identical carrier frequency, optical polarization, and exponential envelope~\cite{Hofer11,Milman17}.	 
	After receiving the results ($X_{1}$, $Y_{2}$),  Bob performs a displacement operation $D(\alpha_{D})$ with $\alpha_{D} \equiv \sqrt{2}(X_{1}-iY_{2})$~\cite{BK98,Opatrny00} onto the magnon mode, which is realized by sending a coherent pulse with duration $\tau_{d}$ to the MW cavity.  

	Again, we assume $\tau_{d}\ll\kappa_{m}^{-1}$ to neglect the magnon dissipation during this displacement pulse. For the resonant case, the corresponding Langevin equations for the averages of the MW cavity and magnon mode are given by~\cite{Lu}
	\begin{equation}
		\begin{aligned}
			&\langle\dot{c}\rangle=-\kappa_{c}\langle c\rangle-ig_{c}\langle m\rangle+E_{d} e^{i \phi},\\
			&\langle\dot{m}\rangle=-ig_{c}\langle c\rangle,
		\end{aligned}
	\end{equation}
	{where $E_{d}=\sqrt{2\kappa_{c}}\langle c^{\rm in}\rangle$ denotes the coupling strength between the cavity and the pulse.} Here, we also introduce the phase $\phi$ of the pulse, which can be adjusted to exactly realize the desired displacement, as shown later.  Since the weak coupling $g_{c}\ll\kappa_{c}$ was assumed in the magnon subtraction operation, after the elimination of the MW cavity, we obtain the differential equation for the magnon average under the displacement pulse
	\begin{equation}
		\langle\dot{m}\rangle=-\mathcal{G}_{c}\langle m\rangle-i \frac{ g_{c}E_{d}e^{i \phi} }{\kappa_{c}}.
	\end{equation}
	Solving the above equation, we get
	\begin{equation}\label{dis}
		\langle m(\tau_{d})\rangle=e^{-\mathcal{G}_{c}\tau_{d}}\langle m(0)\rangle+ i\left(e^{-\mathcal{G}_{c}\tau_{d}}-1\right) \dfrac{E_{d}e^{i \phi} }{g_{c}},
	\end{equation}
	where $\langle m(0)\rangle$ denotes the average of the magnon mode before sending the displacement pulse. Equation~\eqref{dis} indicates that the desired displacement operation is realized when 
	\begin{equation}
		\alpha_{D}  \equiv  \sqrt{2}(X_{1}-iY_{2}) =  i\left(e^{-\mathcal{G}_{c}\tau_{d}}-1\right) \dfrac{E_{d}e^{i \phi} }{g_{c}},
	\end{equation}
	which can be achieved by adjusting the strength and phase of the pulse. Note that Eq.~\eqref{dis} also implies that the average of the magnon mode is slightly reduced due to the displacement pulse~\cite{Milman17}, as seen from $\langle m(0)\rangle\rightarrow\langle m(0)\rangle e^{-\mathcal{G}_{c}\tau_{d}}$. 

	Once Bob completed his displacement operation, the magnon mode is prepared in a state $\rho_{\rm tel}$ that is teleported from the initial optical state $\rho_{\rm in}$. In the formalism of the characteristic function, the teleported magnon state can be expressed as a product of the initial optical state and the shared entangled state~\cite{CF06}
	\begin{equation}
		\chi_{\rm tel}(\alpha)=\chi_{\rm in}(\alpha) \, \chi_{m,1}(\alpha^{*},\gamma\alpha),
	\end{equation}
	where $\alpha$ is a complex variable and $\gamma$ is defined as $\gamma=e^{-\mathcal{G}_{c}\tau_{d}}$. Specifically, the characteristic function of the shared non-Gaussian entangled state is in the form
	\begin{widetext}
		\begin{equation}
			\chi_{m,1}(\alpha^{*},\gamma\alpha)=\left(\dfrac{\lambda'\big(1+\gamma^{2}+3\lambda'^{2}-6\lambda'\gamma+3\lambda'^{2}\gamma^{2}-2\lambda'^{3}\gamma\big)|\alpha|^{2}}{\big(1-\lambda'^{4}\big)}+\dfrac{\lambda'^{2}\big(\lambda'-\gamma\big)^{2}\big(1-\lambda'\gamma\big)^{2}|\alpha|^{4}}{\big(1+\lambda'^{2}\big)\big(1-\lambda'^{2}\big)^{2}}+1\right) \times \exp \left( \dfrac{\big(1-\lambda'\gamma\big)\big(\lambda'-\gamma\big)|\alpha|^{2}}{1-\lambda'^{2}}\right),
		\end{equation}
	\end{widetext}
	where $\lambda'=\tanh r\cos\theta$. To verify the success of our optomagnonic quantum teleportation, we define the fidelity $F$ as the overlap between the initial optical state $\rho_{\rm in}$ and the teleported magnon state $\rho_{\rm tel}$, which is~\cite{DG02}
	\begin{equation}
		F=\dfrac{1}{\pi}\int d^{2}\alpha\,\chi_{\rm in}(\alpha)\chi_{\rm tel}(-\alpha).
	\end{equation}
	The fideity $F$ greater than $0.5$ signifies a successful CV quantum teleportation~\cite{Furu98}.

\begin{figure}[b]
		\includegraphics[width=\linewidth]{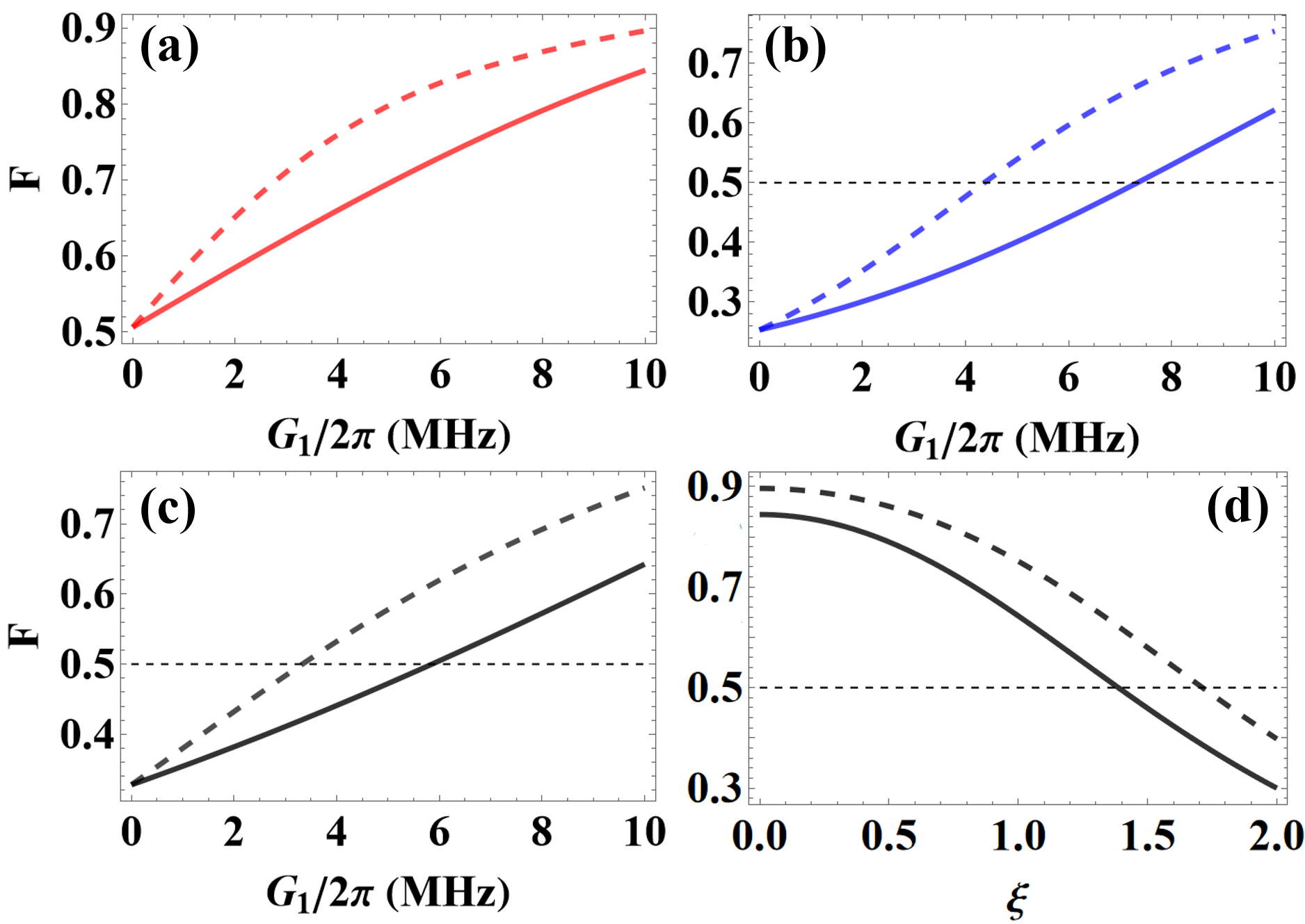}
		\caption{Fidelity $F$ versus the effective optomagnonic coupling strength $G_{1}$ with the shared entangled state being the TMSV state (solid line) or the distilled non-Gaussian state (dashed line) for an initial state being (a) a coherent state $|\beta\rangle$, (b) a single-photon state $|1\rangle$, and (c) a squeezed vacuum state $|\xi\rangle$. (d) Fidelity $F$ versus the squeezing $\xi$ for an initial squeezed state. We take $\xi=1$ in (c), $G_{1}/2\pi=10$ MHz in (d), and $\tau_{d}=10$ ns. The other parameters are the same as in Fig.~\ref{fig2}.}
		\label{fig3}
	\end{figure}

	Below we present the results of the teleportation fidelity for a series of optical input states, including coherent, single-photon, squeezed, and cat states. 	
	For the coherent state $|\beta\rangle$, after some calculation, we obtain the fidelity for the shared entangled state being a TMSV state, given by
	\begin{equation}
		F=\dfrac{2-2\lambda^{2}}{3+\gamma^{2}-4\lambda\gamma-\lambda^{2}\big(1-\gamma^{2}\big)},
	\end{equation}
	where we define $\lambda=\tanh r$. By contrast, for the shared non-Gaussian entangled state, we achieve
	\begin{equation}
		F=\dfrac{\big(1-\lambda'^{2}\big)^{3}\big[(1+\gamma)^{2}-\lambda'(1+\gamma)(1+\gamma^{2})+\lambda'^{2}(1+\gamma^{2})\big]}{(1+\lambda'^{2})\big[1+\gamma-\lambda'(1+\gamma^{2})-\lambda'^{2}(1-\gamma)\big]^{3}}.
	\end{equation}
	In the limit of $\gamma\to1$, i.e., negligible decay of the first term in Eq.~\eqref{dis}, it becomes
	\begin{equation}
		F=\dfrac{\big(1+\lambda'\big)^{3}\big(\lambda'^{2}-2\lambda'+2\big)}{4\big(1+\lambda'^{2}\big)},
	\end{equation}
	which reproduces the result of the CV quantum teleportation in photonic systems~\cite{Kitagawa06,Yang09}.
	In Fig.~\ref{fig3}(a), we show the fidelity $F$ versus the optomagnonic coupling strength $G_{1}$ for an initial coherent state and the entangled state being the TMSV state or the distilled non-Gaussian state.  It shows that the teleportation fidelity increases rapidly with the coupling $G_{1}$, and in the full range of $G_{1}$ the fidelity with the distilled entangled state is always higher than that with the Gaussian entanglement. This clearly confirms that the performance of the CV teleportation is greatly improved by entanglement distillation. Note that the fidelity is in the full range greater than $0.5$, meaning that quantum teleportation of a coherent state is achieved.

	\begin{figure*}[t]
		\hskip-0.5cm \includegraphics[width=\linewidth]{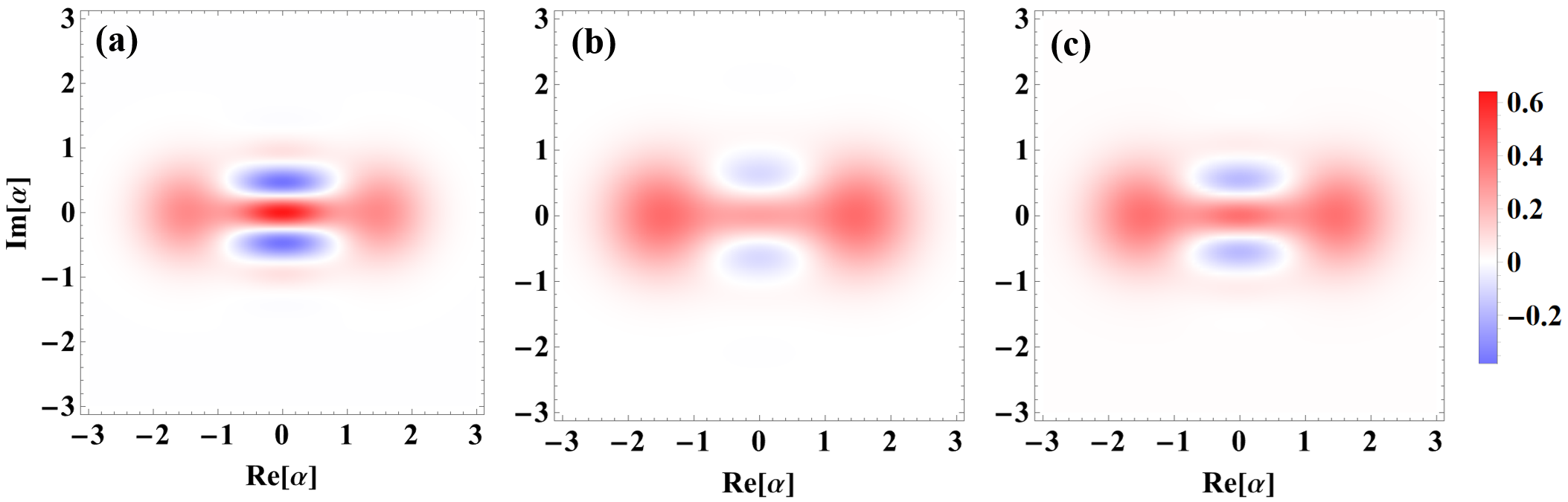}
		\caption{Wigner function of (a) the initial cat state, and of the teleported magnon state by sharing (b) the TMSV state or (c) the distilled non-Gaussian state. We take $\alpha_{0}=1.5$, $\varphi=0$, and $G_{1}/2\pi=10$ MHz. The other parameters are the same as in Fig.~\ref{fig3}.}
		\label{fig4}
	\end{figure*}

	\begin{figure}[h]
		\hskip-0.5cm \includegraphics[width=0.9\linewidth]{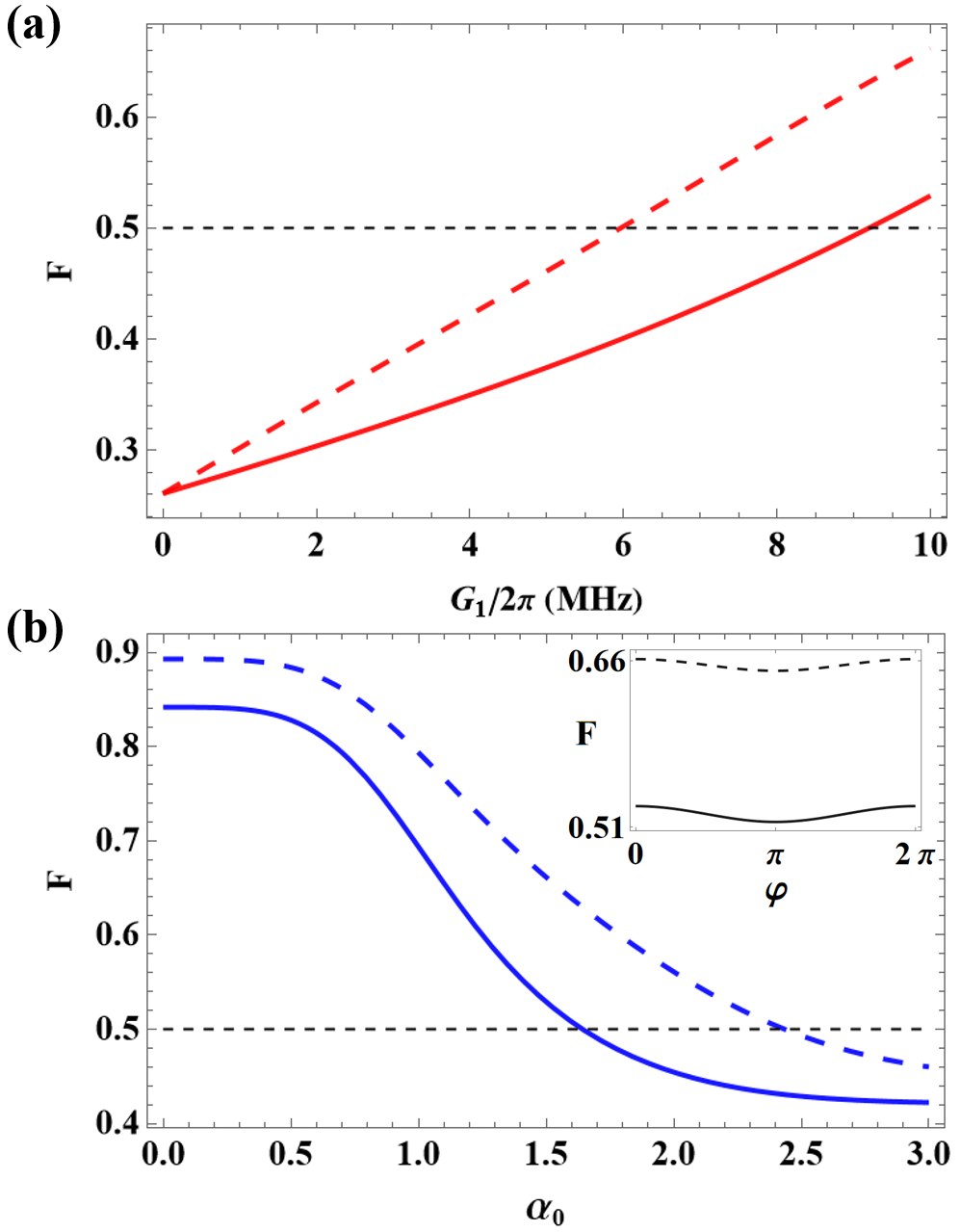}
		\caption{Fidelity $F$ exploiting the TMSV state (solid line) or non-Gaussian entangled state (dashed line) versus (a) the optomagnonic coupling strength $G_{1}$ and (b) the amplitude $\alpha_{0}$ (the phase $\varphi$ in the inset) for an initial cat state. The other parameters are the same as in Fig.~\ref{fig4}.}
		\label{fig5}
	\end{figure}

	For an initial single-photon state $|1\rangle$ or squeezed vacuum state $|\xi\rangle$, the analytical expressions of the fidelity are too lengthy to be reported here but provided in the Appendix. In Fig.~\ref{fig3}(b), we plot the fidelity for the two cases of the shared entangled state. Again, it shows that the entanglement distillation greatly improves the fidelity of the teleportation and reduces the threshold of the optomagnonic coupling strength required for achieving a quantum teleportation ($F>0.5$). A similar finding is observed in Fig.~\ref{fig3}(c) for an initial squeezed state $|\xi\rangle$.  For simplicity, we consider $\xi$ to be real and take $\xi=1$ in Fig.~\ref{fig3}(c). In Fig.~\ref{fig3}(d), we show the fidelity versus the squeezing $\xi$. It tells that the fidelity decreases with $\xi$, suggesting that a smaller squeezing should be used in order to get a higher fidelity~\cite{Anno07,Yang09}.

	Lastly, we consider an initial cat state $|\alpha_{0}\rangle+e^{i\varphi}|-\alpha_{0}\rangle$ (unnormalized). For simplicity, we assume the amplitude $\alpha_{0}$ to be real and introduce the relative phase between the two superposed coherent states. In Fig.~\ref{fig4}, we plot the Wigner function of the initial optical state and the teleported magnon state with the shared TMSV state or distilled non-Gaussian state. Although both the magnon cat states in Fig.~\ref{fig4}(b)-(c) show unambiguous negativity and interference fringes, the teleported state with entanglement distillation is of higher fidelity to the initial state. 
	Since the analytical expression is too cumbersome, we only numerically plot the fidelity in Fig.~\ref{fig5}(a). Compared to other initial states shown above, the cat state requires a larger optomagnonic coupling to achieve a quantum teleportation, making the distillation operation more necessary.  
	In Fig.~\ref{fig5}(b), we also show the fidelity versus the amplitude $\alpha_{0}$ (the phase $\varphi$ in the inset). Clearly, the fidelity reduces with the amplitude $\alpha_{0}$. In contrast with a rapid drop below the threshold $F=0.5$ in the case with the TMSV state, the fidelity achieved with the distilled non-Gaussian state shows a smoother decline. In both cases, the relative phase $\varphi$ of the cat state shows only minor influence on the fidelity, as seen from the inset of Fig.~\ref{fig5}(b).

	\section{Discussion and Conclusion}\label{V}

	A complete teleportation should also include the state readout of the teleported magnon state. This can be realized by the aid of the MW cavity, as suggested in other quantum protocols~\cite{Lu,Jie21X}. To avoid repetitive calculations, we briefly discuss as follows. The magnon state can be read out by sending a weak MW pulse with duration $\tau_{r}$ into the MW cavity and the linear cavity-magnon beam-splitter (state-swap) interaction maps the magnon state to the cavity output field, from which the state can be reconstructed by performing a microwave tomography. 
	
	{In the above demonstration, we have used the magnon dissipation rate $\kappa_{m}/2\pi=0.5$ MHz~\cite{Shen25} for a YIG sphere. This dissipation rate, though small in typical cavity magnonic experiments, still corresponds to a short magnonic coherence time which restricts the maximum communication distance of our protocol. Nonetheless, we remark that  this is a more technical issue and can be significantly improved by increasing the purity of the YIG. For example, a recent experiment has reported an ultra-long magnon lifetime $\sim$18 $\mu$s using an ultra-pure single-crystal YIG sphere~\cite{Bozhko}}.
	
	In conclusion, we have presented an optomagnonic CV quantum teleportation protocol based on a YIG sphere that is placed inside a MW cavity. By using appropriately chosen optical and microwave pulses, we have shown that a series of key operations, including the generation and distillation of the optomagnonic entanglement, the magnon displacement, and the readout of the magnon state, can be accomplished.  Our work extends the CV quantum teleportation to the optomagnonic system and lays the foundation for realizing quantum repeaters and quantum networks based on magnons. The work also offers a route to prepare diverse magnonic quantum states by transferring quantum states from optics to magnonics.

	\section*{ACKNOWLEDGMENTS}
	This work was supported by Zhejiang Provincial Natural Science Foundation of China (Grant No. LR25A050001), National Natural Science Foundation of China (Grant No. 12474365, 92265202) and National Key Research and Development Program of China (Grant No. 2024YFA1408900, 2022YFA1405200).

	\section*{APPENDIX}\label{app}
	\setcounter{figure}{0}
	\setcounter{equation}{0}
	\setcounter{table}{0}
	\renewcommand\theequation{A\arabic{equation}}
	\renewcommand\thefigure{A\arabic{figure}}
	\renewcommand\thetable{A\arabic{table}}
	
	Here, we provide the analytical expression of the fidelity of the teleportation for the input optical state being a single-photon state $|1\rangle$ or a squeezed vacuum state $|\xi\rangle$.
	For the single-photon state, the fidelity with the shared optomagnonic entanglement being a TMSV state is given by
	\begin{equation}
		F=\dfrac{2\big(1-\lambda^{2}\big)\big(a_{0}+a_{1}\lambda+a_{2}\lambda^{2}+a_{3}\lambda^{3}+a_{4}\lambda^{4}\big)}{\big(3+\gamma^{2}-4\lambda\gamma-\lambda^{2}+\lambda^{2}\gamma^{2}\big)},
	\end{equation} 
	where the coefficients are
	\begin{equation}
		\begin{aligned}
			&a_{0}=5+2\gamma^{2}+\gamma^{4},\\
			&a_{1}=-8\gamma-8\gamma^{3},\\
			&a_{2}=-6+20\gamma^{2}+2\gamma^{4},\\
			&a_{3}=-8\gamma-8\gamma^{3},\\
			&a_{4}=5+2\gamma^{2}+\gamma^{4}.
		\end{aligned}
	\end{equation}
	For the shared entanglement being a non-Gaussian entangled state, the fidelity is
	\begin{equation}
		F=\dfrac{\big(b_{0}+b_{1}\lambda'+b_{2}\lambda'^{2}+b_{3}\lambda'^{3}+b_{4}\lambda'^{4}+b_{5}\lambda'^{5}+b_{6}\lambda'^{6}\big)}{\big(1+\lambda'^{2}\big)\big(1-\lambda'^{2}\big)^{-3}\big[1+\gamma-\lambda'\big(1+\gamma^{2}\big)-\lambda'^{2}\big(1-\gamma\big)\big]},
	\end{equation} 
	where the coefficients are
	\begin{equation}
		\begin{aligned}
			&b_{0}=1+2\gamma+2\gamma^{2}+2\gamma^{3}+\gamma^{4},\\
			&b_{1}=1-3\gamma-6\gamma^{2}-6\gamma^{3}-7\gamma^{4}-3\gamma^{5},\\
			&b_{2}=-8\gamma+19\gamma^{2}+24\gamma^{3}+10\gamma^{4}+8\gamma^{5}+3\gamma^{6},\\
			&b_{3}=3-27\gamma-21\gamma^{2}-35\gamma^{3}-27\gamma^{4}-9\gamma^{5}-3\gamma^{6}-\gamma^{7},\\
			&b_{4}=12+20\gamma+55\gamma^{2}+30\gamma^{3}+22\gamma^{4}+10\gamma^{5}+3\gamma^{6},\\
			&b_{5}=-7-27\gamma-18\gamma^{2}-30\gamma^{3}-11\gamma^{4}-3\gamma^{5},\\
			&b_{6}=1+4\gamma+14\gamma^{2}+4\gamma^{3}+\gamma^{4}.
		\end{aligned}
	\end{equation}
	
	For the input squeezed vacuum state, the fidelity with a shared TMSV state is in the form
	\begin{equation}
		F=\big(1+2c_{0}\cosh 2\xi+c_{0}^2\big)^{-\frac{1}{2}},
	\end{equation}
	where $c_{0}$ is given by
	\begin{equation}
		c_{0}=\dfrac{1+\gamma^{2}-4\lambda\gamma+\lambda^{2}\big(1+\gamma^{2}\big)}{2-2\lambda^{2}}.
	\end{equation}
	For the case where Alice and Bob share a distilled non-Gaussian state, the fidelity is obtained as
	\begin{equation}
		F=\dfrac{\big(d_{0}+d_{1}\lambda'+d_{2}\lambda'^{2}+d_{3}\lambda'^{3}+d_{4}\lambda'^{4}+d_{5}\lambda'^{5}+d_{6}\lambda'^{6}\big)}{2\big(1+\lambda'^{2}\big)\big(1-\lambda'^{2}\big)^{-3}\big(d_{7}d_{8}\big)^{\frac{5}{2}}}
	\end{equation}
	with the coefficients
	\begin{equation}
		\begin{aligned}
			&d_{0}=2+8\gamma^{2}+2\gamma^{4}+8\big(\gamma+\gamma^{3}\big)\cosh 2\xi+4\gamma^{2}\cosh 4\xi,\\
			&d_{1}=-12\gamma-18\gamma^{3}-6\gamma^{5}-6\big(1+4\gamma^{2}+3\gamma^{4}\big)\cosh 2\xi\\
			&\qquad-6\big(\gamma+\gamma^{3}\big)\cosh 4\xi,\\
			&d_{2}=2+15\gamma^{2}+22\gamma^{4}+6\gamma^{6}+8\big(2\gamma+9\gamma^{3}+3\gamma^{5}\big)\cosh 2\xi\\
			&\qquad+\big(2+7\gamma+2\gamma^{4}\big)\cosh 4\xi,\\
			&d_{3}=4\gamma-12\gamma^{3}-18\gamma^{5}-2\gamma^{7}+2\big(1-7\gamma^{2}-9\gamma^{4}-\gamma^{6}\big)\cosh 2\xi\\
			&\qquad-2\big(\gamma+\gamma^{3}\big)\cosh 4\xi,\\
			&d_{4}=-5-6\gamma^{2}+15\gamma^{4}+6\gamma^{6}+\big(1+4\gamma^{2}+\gamma^{4}\big)\cosh 4\xi,\\
			&d_{5}=2\gamma-4\gamma^{3}-6\gamma^{5}+2\big(1+4\gamma^{2}+3\gamma^{4}\big)\cosh 2\xi\\
			&\qquad-4\big(\gamma+\gamma^{3}\big)\cosh 4\xi,\\
			&d_{6}=2+\gamma^{2}+2\gamma^{4}-4\big(\gamma+\gamma^{3}\big)\cosh 2\xi+3\gamma^{2}\cosh 4\xi,\\
			&d_{7}=\gamma-\lambda'-\lambda'\gamma^{2}+\lambda'^{2}\gamma+\big(1-\lambda'^{2}\big)e^{2\xi},\\
			&d_{8}=\gamma-\lambda'-\lambda'\gamma^{2}+\lambda'^{2}\gamma+\big(1-\lambda'^{2}\big)e^{-2\xi}.
		\end{aligned}
	\end{equation}

\end{document}